\documentclass[10pt,aps,pra,groupedaddress,showpacs,twocolumn,showkeys]{revtex4-1}

\usepackage{amsmath,amssymb,graphicx}

\begin{document}

\title{Equalizing the near and far electromagnetic fields around particles made of different materials}
\author{Alexis Devilez}
\email{alexis.devilez@fresnel.fr}
\affiliation{CNRS, Aix-Marseille Universit\'e, Centrale Marseille, Institut Fresnel, UMR 7249, Campus de St J\'er\^ome, 13397 Marseille, France}
\author{Xavier Zambrana-Puyalto}
\affiliation{CNRS, Aix-Marseille Universit\'e, Centrale Marseille, Institut Fresnel, UMR 7249, Campus de St J\'er\^ome, 13397 Marseille, France}
\author{Brian Stout}
\affiliation{CNRS, Aix-Marseille Universit\'e, Centrale Marseille, Institut Fresnel, UMR 7249, Campus de St J\'er\^ome, 13397 Marseille, France}
\author{Nicolas Bonod}
\email{nicolas.bonod@fresnel.fr}
\affiliation{CNRS, Aix-Marseille Universit\'e, Centrale Marseille, Institut Fresnel, UMR 7249, Campus de St J\'er\^ome, 13397 Marseille, France}

\begin{abstract}
We demonstrate that the electromagnetic fields scattered by particles made of different materials can be equalized. Emphasize is placed first in metallic nanoparticles that host localized surface plasmons and it is shown that their electromagnetic fields can be identically reproduced with dielectric particles. We derive the explicit formulas relating the different constitutive parameters that yield identical electromagnetic responses. This method provides the dielectric permittivities of spherical particles that reproduce the strong near electric field intensities observed around metallic particles featuring localized surface plasmon resonances in optics or near infrared frequencies. We also demonstrate the ability of homogenous dielectric particles to host the magnetic resonances predicted for exotic materials with negative permeability. 
\end{abstract}


\maketitle 
Localized surface plasmon resonances on metallic nanoparticles arise from the resonant oscillation of electrons in the conduction band \cite{Murray2007,Maier2007,Enoch2012}. Plasmonic nanoparticles have received  growing attention over the last 15 years due to their ability to strongly enhance the light matter interaction at a nanometer scale with groundbreaking innovations spanning from cancer thermotherapy \cite{Bardhan2011} or solar cells \cite{Atwater2010} to optical antennas \cite{Novotny11} and enhanced spectroscopy \cite{Yampolsky2014}. Noble metals have attracted most of the efforts to design plasmonic particles, but alternative materials such as doped semi-conductors with tunable free carrier concentrations or transition metal nitrides offer a high potential to create robust and cost effective plasmonic particles \cite{Luther2011,Naik2011,Garcia2011,Guler2015}.

On the other hand, dielectric materials do not have electrons in the conduction band and their electromagnetic properties are characterized by their bound electrons \cite{Webb2011}. 
However, it is now well known that dielectric particles also support electromagnetic resonances that are commonly called Mie or morphologic resonances \cite{Bohren1983,Schuller07,Zhao09}. Interestingly, Mie resonances lead to strong electric or magnetic responses and they can be used to enhance either the electric or the magnetic field intensity \cite{GarciaEtxarri11, Zambrana2012, Shi13,Albella2013,Zywietz2014,Boudarham2014}. Bridging the gap between the electromagnetic modes hosted by particles filled by materials with or without free electrons is of crucial importance to design cost effective and versatile resonant particles with \textit{ad-hoc} properties throughout the relevant spectral domains.

In this work, we show that the electromagnetic (EM) response of plasmonic nanoparticles can be identically reproduced by homogeneous dielectric particles of the same size parameter $z_{i}$ ($z_{i}=k_{i}R$ with $k_{i} ^2= \varepsilon_i\mu_i k_0^2$, $k_0$ being the wavenumber in vacuum, $R$ the radius of the particle, $\varepsilon_i$ the dielectric permittivity and $\mu_i$ magnetic permeability of the particle). The demonstration is carried out for particles much smaller than the wavelength, $\lambda$, such that their EM properties can be described by their first order electric and magnetic coefficients, $a_{1}$ and $b_{1}$ only \cite{Bohren1983,Mishchenko2002, Stout08}.  
First, we show that the EM responses of dielectric and metallic particles of the same size can be equivalent, $i.e.$ their scattering response to a given incident field can be strictly identical. In this case, the two particles respond with the same complex electric dipolar polarizability $\alpha_e$, \textit{i.e.} the same Mie coefficients since $\alpha_e =-\frac{6\pi }{ik^{3}}a_{1}$ \cite{Born00}. 
This method is then extended to particles exhibiting magnetic permeability with a particular interest given to exotic particles of negative permeability. Such particles are interesting because they support a magnetic mode due to a pole of the magnetic Mie coefficient $b_{1}$, similarly to the case of metallic particles with negative permittivity for which the localized plasmon resonance are characterized by a pole of the electric Mie coefficient $a_{1}$. We demonstrate that the electromagnetic mode of the magnetic Mie coefficient $b_{1}$ can be identically obtained with homogeneous dielectric particles and can lead to strong magnetic near-field intensities. 

The electromagnetic response of spherical particles can be classically described by the T-matrix linking the outgoing (Hankel) fields with the incident (Bessel) fields \cite{Grigoriev2013,Grigoriev2015}. The electric elements of this matrix at the $n^{th}$ multipolar order can be written:
\begin{equation}
a_{n}=\frac{j_{n}(z_{b})}{h_{n}^{(+)}(z_{b})}\frac{ \varepsilon
_{s}\varphi _{n}^{(1)}(z_{b})-\varepsilon _{b}\varphi _{n}^{(1)}(z_{s})%
}{ \varepsilon _{s}\varphi _{n}^{(+)}(z_{b})-\varepsilon
_{b}\varphi _{n}^{(1)}(z_{s})}.
\label{an}
\end{equation}
The background and particle dielectric permittivities are respectively denoted $\varepsilon_b$ and $\varepsilon_s$. In what follows, the external host medium is assumed to be the vacuum, $i.e.$ $\varepsilon_b = 1$. The spherical Bessel and Hankel functions are respectively denoted, $j_{n}$ and $h_{n}^{\pm}$, the $\pm$ denoting the outgoing (+) and incoming (-) Hankel functions. The $\varphi_n$ functions are defined as:
\begin{eqnarray}
\varphi _{n}^{(\pm)}(z) &\equiv &\frac{\left[ zh_{n}^{(\pm)}(z)\right] ^{\prime }%
}{h_{n}^{(\pm)}(z)}, \\
\varphi _{n}^{(1)}(z) &\equiv &\frac{\left[ zj_{n}(z)\right] ^{\prime }}{%
j_{n}(z)}.
\label{phin+}
\end{eqnarray}

We first address the case of particles featuring electric dipolar resonances classically described by the Mie coefficient $a_1$. The question is whether particles with the same size parameter but composed of two different materials can feature equivalent dipolar scattering responses, {\it i.e.} \textit{via} $a_{1}(z_0,\varepsilon _{\rm in},\mu_{\rm in})=a_{1}(z_0,\varepsilon_{\rm eq},\mu_{\rm eq})$ where the subscripts "in" and "eq" denote respectively the initial material and the equivalent counterpart. This relation can be written using Eq.~\ref{an} as:
\begin{equation}
\frac{ \varepsilon _{\rm in}\varphi _{1}^{(1)}(z_{0})-\varphi
_{1}^{(1)}(z_{\rm in}) }{ \varepsilon _{\rm in}\varphi
_{1}^{(+)}(z_{0})-\varphi _{1}^{(1)}(z_{\rm in}) }=\frac{
\varepsilon _{\rm eq}\varphi _{1}^{(1)}(z_{0})-\varphi _{1}^{(1)}(z_{\rm eq}) 
}{ \varepsilon _{\rm eq}\varphi _{1}^{(+)}(z_{0})-\varphi
_{1}^{(1)}(z_{\rm eq}) }
\end{equation}
that leads to the following transcendental equation:
\begin{equation}
\frac{\varphi_{1}(z_{\rm in})}{\varepsilon _{\rm in}}=
\frac{\varphi_{1}(z_{\rm eq})}{\varepsilon _{\rm eq}}.
\label{equiv_a_gen}
\end{equation}

We propose to solve this equation by applying the Weierstrass factorization \cite{Grigoriev2013} to the Bessel functions, $j_{n}$ which gives the following expansion for $\varphi_n$: \cite{Grigoriev2015}
\begin{equation}
\varphi _{n}^{(1)}(z)=n+1+\sum_{\alpha =1}^{\infty }\frac{2z^{2}}{z^{2}-a_{n,\alpha }^{2}},
\end{equation}
In the latter expression, $a_{n,\alpha }$ are the (tabulated) zeros of the Bessel function $j_{n}(x)$. 
We assume that $\varphi_1$ can be appropriately approximated for small particles by taking only the lowest zero, $a_{1,1}$, leading to:

\begin{equation}
\varphi _{1}^{(1)}(z)\simeq 2\frac{1-(z/b)^{2}}{1-(z/a)^{2}}, 
\label{phinweier}
\end{equation}
with $a = a_{1,1} =4.493$ and $b=a_{1,1}/\sqrt{2}=3.177$. 

\begin{widetext}
\begin{figure*}[!htb]
\begin{center}
\includegraphics[width=0.65\linewidth]{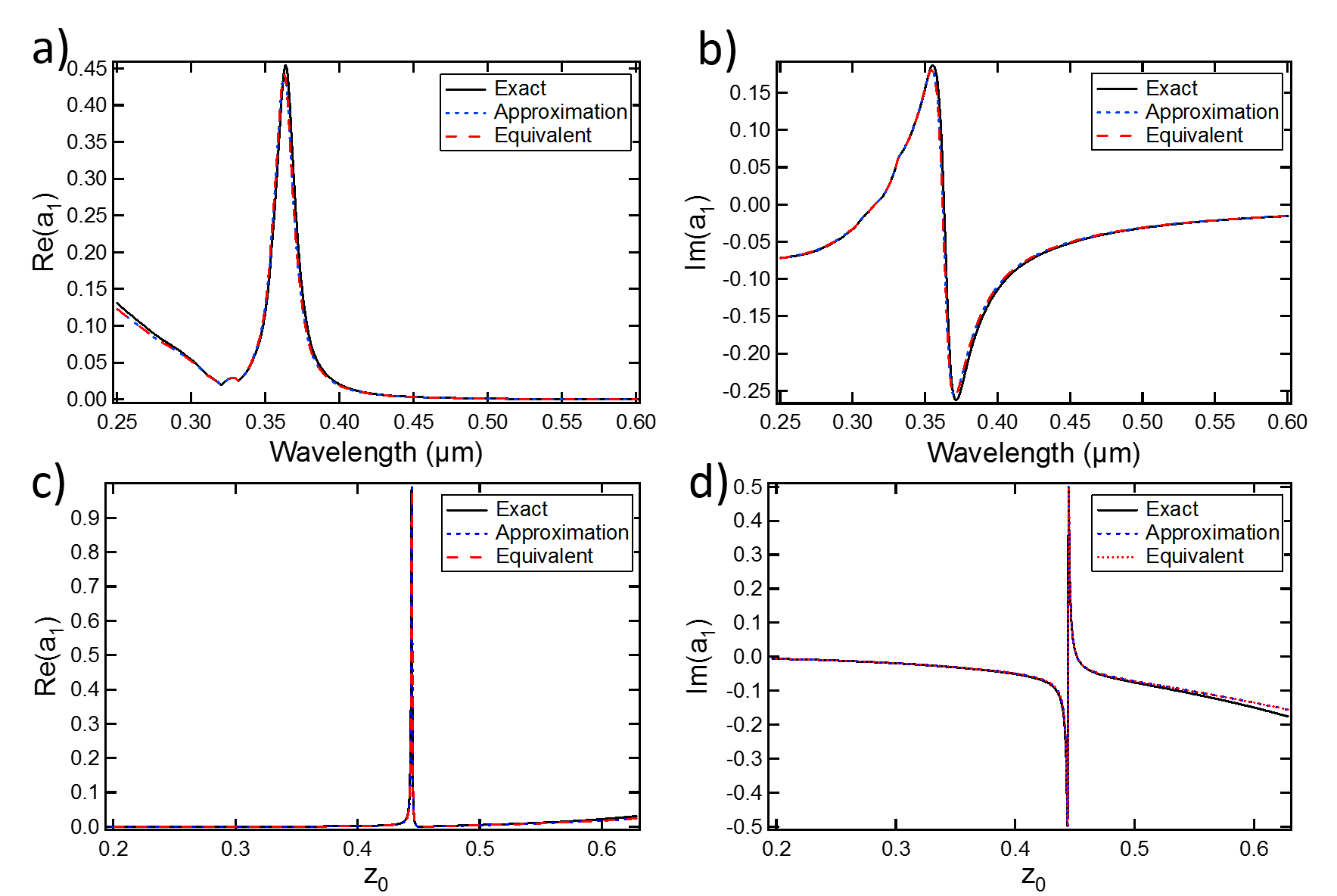} 
\end{center}
\caption {(Color online). Plots of the dipolar Mie coefficient $a_1$ for metallic and dielectric particles. The exact values of $a_1$ are plotted in black full lines, the approximation value using Eq.~\ref{phinweier} is plotted in blue dotted lines and the coefficient $a_1$ of the equivalent permittivity obtained from Eq.~\ref{eps1} is displayed with red dashed lines. a) Real and b) imaginary parts for a silver particle, 50 nm in diameter, with respect to the wavelength. c) Real and d) imaginary parts for a dielectric particle of permittivity $\varepsilon_{\rm in}$ = 100 with respect to the size parameter $z_{i}=k_{i}R$. The size parameter range is taken equal to that with the metallic particle.}
\label{a1_approx}
\end{figure*}
\end{widetext}

It should be stressed that by taking a single pole, some solutions are omitted. However these solutions generally lead to higher permittivities than the first order solution so they are ignored.
The ability of this approximation to accurately describe the resonance for both metallic and dielectric particles is illustrated in Fig.~\ref{a1_approx}. 
The exact and approximated real parts of the dipolar Mie coefficient $a_1$ are displayed in Fig.~\ref{a1_approx}(a-b) as a function of the wavelength when considering a silver particle ($\varepsilon_{\rm in}$ = $\varepsilon_{\rm Ag}$) of diameter $\phi$= 50 nm. They are also plotted in the case of dielectric particles in Fig.~\ref{a1_approx}(c-d) with $\varepsilon_{\rm in}$ = 100. Such dielectric permittivities being observed in other spectral domains, they are plotted with respect to the size parameter $z_{i}=k_{i}R$. It can be observed that the polarizability of both dielectric and metallic particles is accurately described by Eq.~\ref{phinweier}. 

This agreement allows us to cast Eq.~\ref{equiv_a_gen} as:
\begin{equation}
\varepsilon _{\rm in}\frac{1-\varepsilon _{\rm eq}(z_{0}/b)^{2}}{1-\varepsilon
_{\rm eq}(z_{0}/a)^{2}}=\varepsilon _{\rm eq}\frac{1-\varepsilon _{\rm in}(z_{0}/b)^{2}}{1-\varepsilon _{\rm in}(z_{0}/a)^{2}},
\label{sym_approx}
\end{equation}
where $\mu_{\rm in}=\mu_{\rm eq}=1$. We remind that $\varepsilon_{\rm in}$ describes the initial material dielectric permittivity that will be mimicked by an equivalent permittivity, $\varepsilon_{\rm eq}$.
Eq.~\ref{sym_approx} leads to a $2^{nd}$ order polynomial expression with respect to $\varepsilon_{\rm in}$ and has two solutions. A trivial solution satisfying $a_{1}(z_{0},\varepsilon_{\rm eq})=a_{1}(z_0,\varepsilon _{\rm in})$ is given by $\varepsilon_{\rm eq}$ = $\varepsilon_{\rm in}$. The non trivial solution $\varepsilon_{\rm eq}$ can be written as:
\begin{eqnarray}
\varepsilon _{\rm eq} &=&\left( \frac{a}{z_{0}}\right) ^{2} \frac{%
1-\varepsilon _{\rm in}(z_{0}/a)^{2}}{1-\varepsilon _{\rm in}(z_{0}/b)^{2}}.
\label{eps1}
\end{eqnarray}

Eq.~\ref{eps1} provides a simple analytical relation to calculate an equivalent permittivity $\varepsilon_{\rm eq}$ so that the particle scattering response in the electric dipolar regime is identical to that of the initial particle.
The $a_1$ coefficient is also displayed in Fig.~\ref{a1_approx} using the equivalent permittivity. It shows an excellent agreement with the initial $a_1$ coefficient demonstrating the relevance of our approach.
\begin{figure}[!htb]
\begin{center}
\includegraphics[width=1\linewidth]{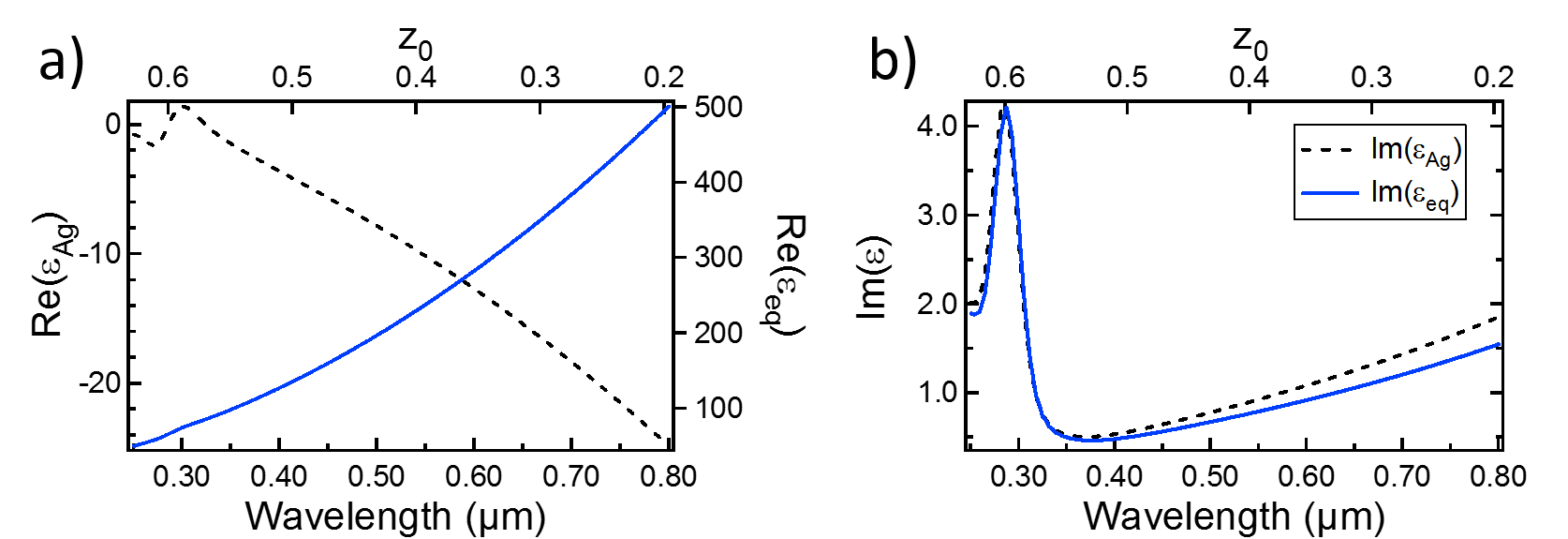} 
\end{center}
\caption {(Color online). Real (a) and imaginary (b) parts of silver permittivity given in \cite{Rakic1998} (black dotted lines), and the dielectric equivalent permittivity (blue full lines) obtained with Eq.~\ref{eps1} as a function of the size parameter (top scale) and the corresponding optical wavelength (bottom scale) calculated in the case of a particle of diameter 50 nm. In a), the scale of the real part of the equivalent permittivity is given on the right side of the graph.}
\label{silverdielec}
\end{figure}

Fig.~\ref{silverdielec} displays the silver permittivity (left scale) as a function of the size parameter (top scale) (it is also displayed with respect to the wavelength with $\phi$=50 nm, together with the equivalent permittivity calculated by using Eq.~\ref{eps1} (right scale). It shows that the resonant scattering properties in the visible spectrum of a spherical particle made of silver with $\phi$=50 nm, can be reproduced by a particle made of a dielectric at a smaller frequency but with the same size parameter, with a real part of the dielectric permittivity ranging from 50 to 500. Such values can be obtained in microwaves which means that the plasmonic properties of metallic nanoparticles can be reproduced with homogeneous dielectric spheres in the hyperfrequency domain.
 
Let us now study the sign of the equivalent dielectric permittivity in order to determine whether both dielectrics and metals can reproduce the optical response of a given dielectric particle. Eq.~\ref{eps1} implies that $\varepsilon _{\rm eq}$ has one pole at $z_{0}=b/\sqrt{\varepsilon _{\rm in}}$ and one zero at $z_{0}=a/\sqrt{\varepsilon_{\rm in}}$ and it can be shown that when $\varepsilon_{\rm in}\in\mathbb{R}^{+}$, $Re(\varepsilon_{\rm eq})<0$ if $a/\sqrt{2}=b<\sqrt{\varepsilon_{\rm in}}z_{0}=z_{\rm in}<a$.
The important conclusion is that the dipolar properties of a dielectric particle can be reproduced by another dielectric if $z_{\rm in} < b$ or $z_{\rm in} > a$, or by a metal if $b<z_{\rm in}<a$. 

\begin{figure}[!htb]
\begin{center}
\includegraphics[width=1\linewidth]{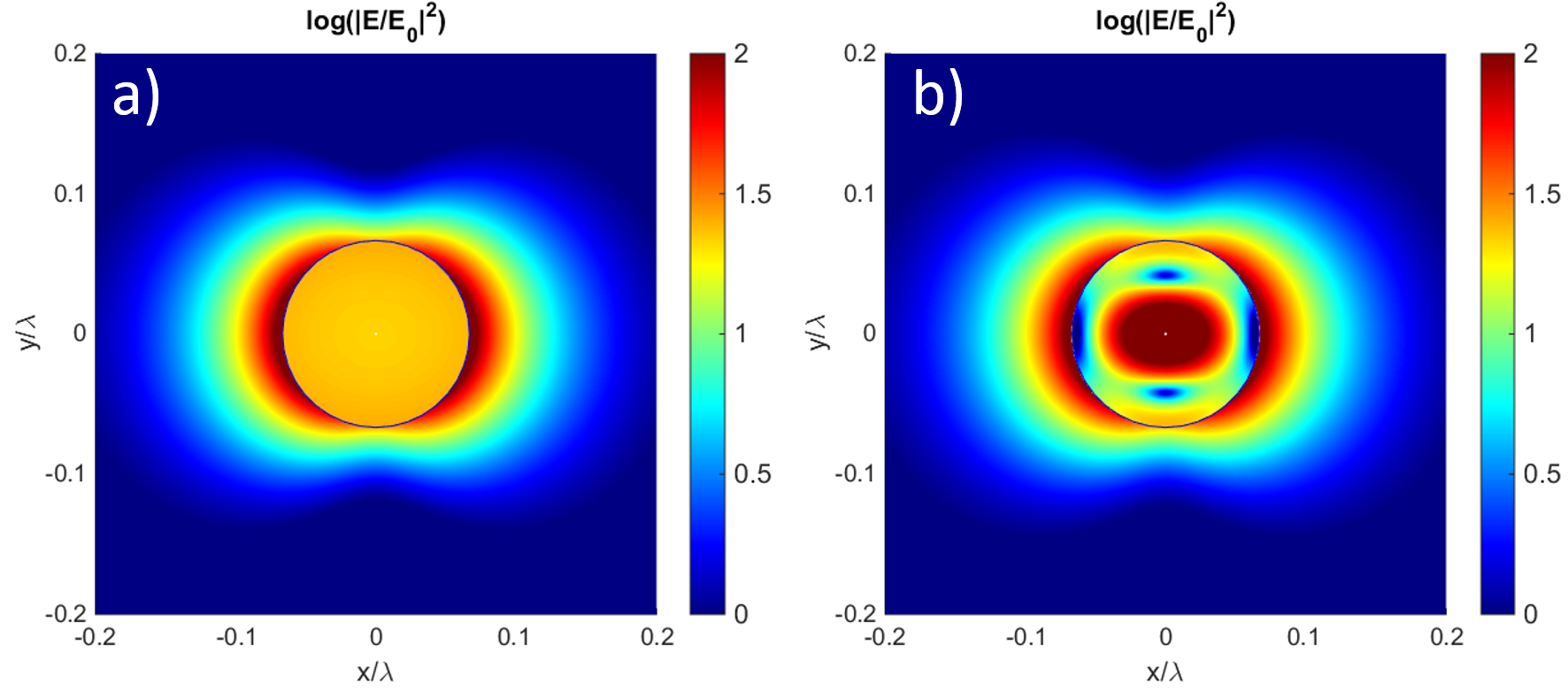} 
\end{center}
\caption {(Color online). The electric field intensity is reconstructed around a particle of diameter $\phi$=50 nm at $\lambda =373.6$ nm, but with a) the dielectric permititivity of silver $\varepsilon_{\rm in}=\varepsilon_{\rm Ag}$ taken from \cite{Rakic1998} and b) with $\varepsilon_{\rm eq}=111.83+i0.47$ obtained with Eq.~\ref{eps1}}
\label{fieldE}
\end{figure}

To further highlight the equivalence of the scattering properties between metals and dielectrics, the total electric field intensity (internal and scattered + incident) is reconstructed in Fig.~\ref{fieldE} around a particle of diameter $\phi$=50 nm for a plane wave illumination at $\lambda =373.6$ nm corresponding to the plasmonic resonance of the silver particle (see Fig.~\ref{a1_approx}a), with the dielectric permittivity of silver $\varepsilon_{\rm in}=\varepsilon_{\rm Ag}$ \cite{Rakic1998} and in Fig.~\ref{fieldE}(a) and with $\varepsilon_{\rm eq}=111.83+i0.47$ obtained with Eq.~\ref{eps1} in Fig.~\ref{fieldE}(b). We observe that the field intensities outside the particles are strictly identical.

A similar methodology can also be applied to prove the equivalence between dielectric and magnetic materials. For that purpose, we consider two particles with the same size parameter. The dipolar magnetic scattering coefficient $b_1$ of a dielectric particle of permittivity, $\varepsilon_{\rm eq}$, is taken to be equal to that of a particle of permeability $\mu_{\rm in}$:
\begin{equation}
b_{1}(z_{0},\varepsilon_{\rm eq})=b_{1}(z_{0},\mu_{\rm in}),
\end{equation}
with $\mu_{\rm eq}=1$ and $\varepsilon_{\rm in}=1$. The magnetic elements $b_n$ at the $n^{th}$ multipolar order are defined as:
\begin{equation}
b_{n}=\frac{j_{n}(z_{b})}{h_{n}^{(+)}(z_{b})}\frac{ \mu _{s}\varphi
_{n}^{(1)}(z_{b})-\mu _{b}\varphi _{n}^{(1)}(z_{s})}{ \mu
_{s}\varphi _{n}^{(+)}(z_{b})-\mu _{b}\varphi _{n}^{(1)}(z_{s})}.
\label{bn}
\end{equation}

This leads to the following transcendental equation:
\begin{equation}
\varphi _{1}^{(1)}(z_{\rm eq})=\frac{\varphi _{1}^{(1)}(z_{\rm in})}{\mu _{\rm in}}.
\label{trans_b1}
\end{equation}

The approximation used in the first part of Eq.~\ref{phinweier} is not accurate enough to solve the transcendental equation of the magnetic response, $b_1$, in particular at the particle resonance. 
To solve Eq.~\ref{trans_b1}, a Newton algorithm can be employed. 
However, Eq.~\ref{mu_app} obtained by plugging Eq.~\ref{phinweier} into Eq.~\ref{trans_b1} is used to provide an initial input value to the algorithm: 
\begin{equation}
+\mu_{\rm in}\frac{1-\varepsilon _{\rm eq}(z_{0}/b)^{2}}{1-\varepsilon
_{\rm eq}(z_{0}/a)^{2}}=\frac{1-\mu_{\rm in}(z_{0}/b)^{2}}{1-\mu_{\rm in}(z_{0}/a)^{2}},
\label{mu_app}
\end{equation}
giving the following approximated solution:
\begin{widetext}
\begin{equation}
\varepsilon _{\rm eq}=-\frac{\left( \mu _{\rm in}-1-\mu _{\rm in}^{2}(z_{0}/a)^{2}+\mu
_{\rm in}(z_{0}/b)^{2}\right) }{\left[ -\mu _{\rm in}(z_{0}/b)^{2}-\mu
_{\rm in}(z_{0}/a)^{2}(z_{0}/b)^{2}+\mu
_{\rm in}^{2}(z_{0}/b)^{2}(z_{0}/a)^{2}+(z_{0}/a)^{2}\right]}.
\label{sol_mu}
\end{equation}
\end{widetext}

\begin{figure}[!htb]
\begin{center}
\includegraphics[width=1\linewidth]{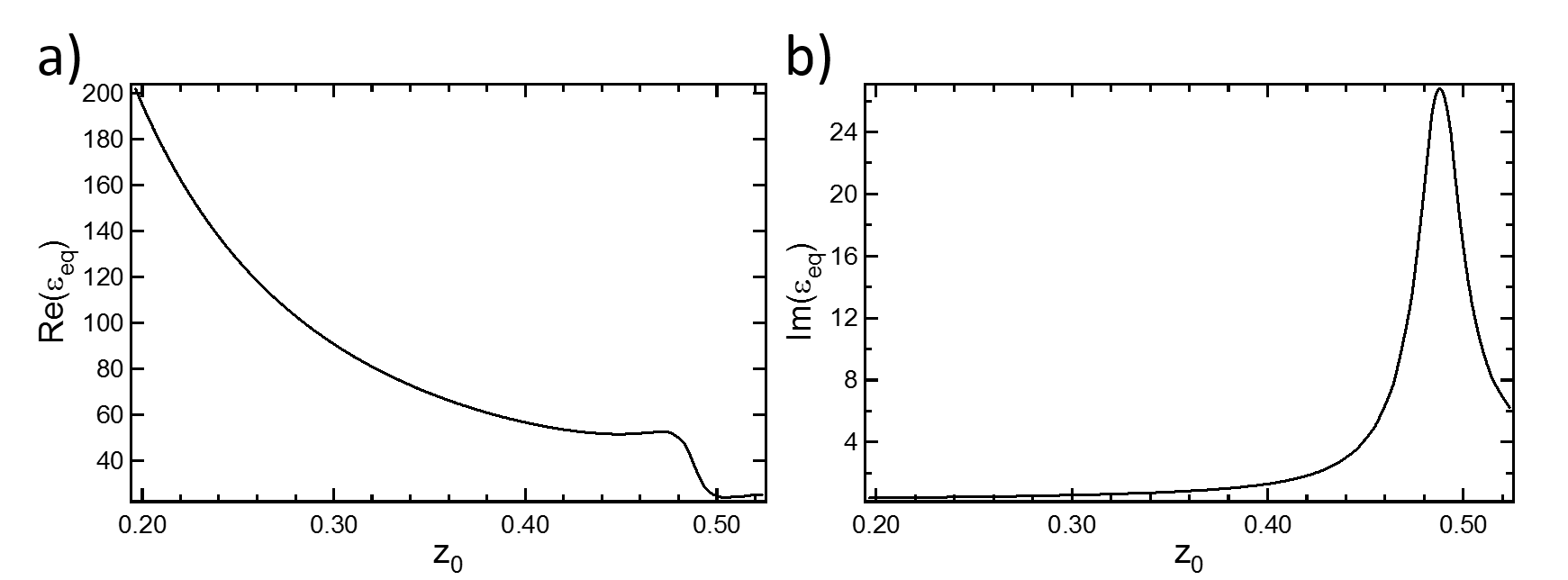} 
\end{center}
\caption {Real (a) and imaginary (b) parts of the dielectric equivalent permittivity $\varepsilon_{\rm eq}$ in the case of a material featuring a negative permeability $\mu_{\rm in}=\varepsilon_{\rm Ag}$ in the optical range of frequencies as a function of the parameter size (the parameter size is identical to that used in Fig.\ref{silverdielec}.}
\label{mueq}
\end{figure}

Let us now study the case of a material featuring a negative permeability, \textit{e.g.} $\mu_{\rm in}=\varepsilon_{\rm Ag}$, and a dielectric permittivity $\varepsilon_{\rm in} = 1$. In this case, the Mie coefficient $b_1$ of this particle will be equal to the Mie coefficient $a_1$ of the spherical particle made of silver. The coefficient $b_1$ will exhibit a resonance at $\lambda=373.6$ nm for a particle of diameter $\phi$=50 nm.

Let us now show that this magnetic resonance can be fully reproduced with a purely dielectric and homogeneous material. Eq.~\ref{trans_b1} is solved numerically in the case of a 50 nm diameter particle with $\varepsilon_{\rm in}=1$ and $\mu_{\rm in}=\varepsilon_{\rm Ag}$ in the optical frequenices. The plots shown in Fig.~\ref{mueq} show that dielectric particles can reproduce the magnetic plasmon-like resonance ($\mu_{\rm eq}=\varepsilon_{\rm Ag}$). The dielectric permittivities leading to a resonant magnetic response ranges between from 20 to 200. These values are smaller than those obtained for the equivalence between metals and dielectrics displayed in Fig.~\ref{silverdielec}, meaning that a larger variety of dielectric materials can be used to enhance the magnetic near-field counterpart of electromagnetic waves. 

\begin{figure}[!htb]
\begin{center}
\includegraphics[width=1\linewidth]{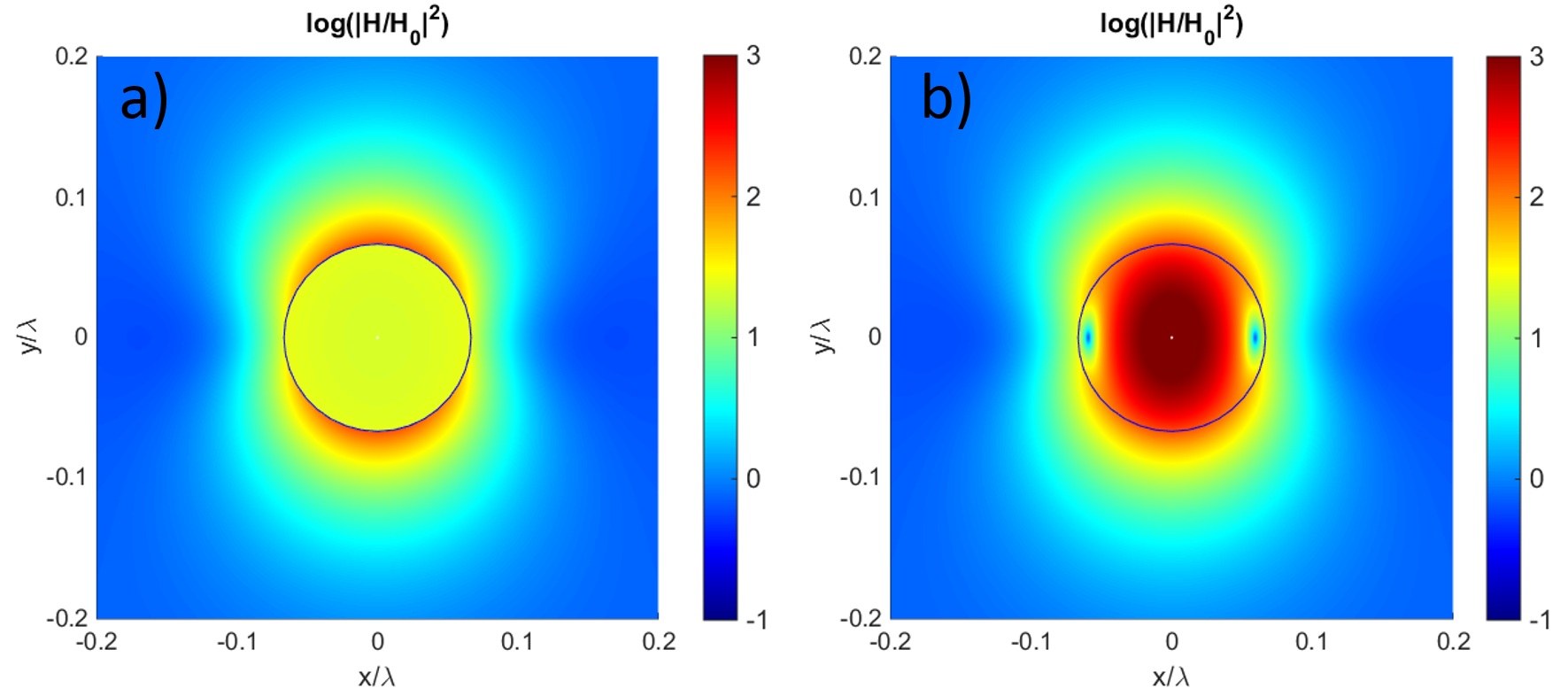} 
\end{center}
\caption {(Color online). Magnetic field intensity around a particle 50 nm in diameter at $\lambda =373.6$ nm, with a) $\mu_{\rm in}= - 2.5+ i0.5$ and b) $\varepsilon_{\rm eq}=53.59+i1.86$.}
\label{fieldH}
\end{figure}

For the sake of completeness, the total magnetic field is displayed in Fig.~\ref{fieldH}a) around a 50 nm particle with $\varepsilon_{\rm in}=1$ and $\mu_{\rm in}=\varepsilon_{\rm Ag}$, at the wavelength $\lambda = 373.6$ nm corresponding to the magnetic dipolar resonance. This resonance is expected to lead to a strong enhancement of the near magnetic field in a manner analogous to the electric plasmon resonance enhancing the near electric field at the vicinity of metallic particles. The near magnetic field is then displayed in Fig.~\ref{fieldH}b) for the equivalent particle, of same size parameter and with a dielectric permittivity $\varepsilon_{\rm eq} = 53.59 + i1.86$ that satisfies Eq.~\ref{trans_b1}. It can be observed that the magnetic fields scattered by the two particles are strictly identical. This result illustrates that homogeneous dielectric particles can be used to reproduce the electromagnetic response of magnetic materials. Let us recall that this equivalence is valid exterior to the particles in the near, intermediate and far fields. We emphasize that the electromagnetic fields differ strongly inside the particles due to the absence of skin depth in dielectric materials.


To conclude, we showed the ability of dielectric particles to feature the same electromagnetic response as particles made with metallic or magnetic materials. The multipolar theory combined with the Weierstrass factorization allowed us to derive the explicit relation between the two dielectric permittivites in the former case, and between the dielectric permittivity and magnetic permeability in the latter case. This equivalence bridges the gap between plasmon resonances observed with metallic nanoparticles in optics and dielectric materials that can be found at smaller frequencies. Importantly, it also permits to reproduce the electromagnetic properties of exotic magnetic materials that feature negative magnetic permeabilities to create strong magnetic near field intensities. The method introduced in this manuscript is quite versatile and can be used to design a wide variety of all-dielectric components.

\section*{Acknowledgements}
This work has been carried out thanks to the support of the A*MIDEX project (n$^{\circ}$ ANR-11-IDEX-0001-02) funded by the Investissements d'Avenir French Government program managed by the French National Research Agency (ANR).

\end{document}